%% file: three_flavour_paper.tex
\documentclass[aps,prl,twocolumn,amsmath,floatfix,superscriptaddress,preprintnumbers,showpacs]{revtex4-1}

\usepackage{lineno}
\usepackage{graphicx}
\usepackage{dcolumn}
\usepackage{bm}

\begin{document}

\pagestyle{plain}

\preprint{FERMILAB-PUB-14-028-E-PPD}


\title{Combined analysis of $\nu_{\mu}$ disappearance and $\nu_{\mu} \rightarrow \nu_{e}$ appearance in MINOS using accelerator and atmospheric neutrinos}


%
\input{ThreeFlavor-Nov13-authors.tex}

\date{\today}

\begin{abstract}

We report on a new analysis of neutrino oscillations in MINOS 
using the complete set of accelerator and atmospheric data.
The analysis combines the $\nu_{\mu}$ disappearance and
$\nu_{e}$ appearance data using the three-flavor formalism.
We measure $|\Delta m^{2}_{32}|=[2.28-2.46]\times10^{-3}\mbox{\,eV}^{2}$ (68\% C.L.)
and $\sin^{2}\theta_{23}=0.35-0.65$ (90\% C.L.) in the normal hierarchy,
and $|\Delta m^{2}_{32}|=[2.32-2.53]\times10^{-3}\mbox{\,eV}^{2}$ (68\% C.L.)
and $\sin^{2}\theta_{23}=0.34-0.67$ (90\% C.L.) in the inverted hierarchy.
The data also constrain $\delta_{CP}$, the $\theta_{23}$ octant degeneracy 
and the mass hierarchy; we disfavor 36\% (11\%) of this three-parameter 
space at 68\% (90\%) C.L.

\end{abstract}

\pacs{14.60.Pq}

\maketitle


The study of neutrino oscillations 
has entered a precision era in which the experimental data can be 
used to probe the three-flavor framework of mixing
between the neutrino flavor eigenstates ($\nu_{e}$, $\nu_{\mu}$, $\nu_{\tau}$)
and mass eigenstates ($\nu_{1}$, $\nu_{2}$, $\nu_{3}$).
In the standard theory, neutrino mixing is described by the 
unitary PMNS matrix~\cite{Maki:1962mu,*Pontecorvo:1967fh,*Gribov:1968kq},
parameterized by three angles 
$\theta_{12}$, $\theta_{23}$, $\theta_{13}$
and a phase $\delta_{CP}$.
The oscillation probabilities additionally depend on the two mass-squared differences
$\Delta m^{2}_{32}$ and $\Delta m^{2}_{21}$, where $\Delta m^{2}_{ij} \equiv m^{2}_{i}-m^{2}_{j}$.
The current generation of experiments 
has measured all three mixing angles and the mass-squared differences
using accelerator, atmospheric, reactor and solar neutrinos~\cite{ThePDG}.
Most recently, the smallest mixing angle, $\theta_{13}$, has been measured precisely
by reactor neutrino experiments~\cite{An:2013zwz,Ahn:2012nd,Abe:2012tg}.
However, the picture is not yet complete.
The value of $\delta_{CP}$, which determines the level of CP violation in the lepton sector,
has not yet been measured.
It is also not known whether the neutrino
mass hierarchy is normal ($\Delta m^{2}_{32}>0$) 
or inverted ($\Delta m^{2}_{32}<0$),
whether $\sin^{2}2\theta_{23}$ is maximal,
or if not, whether the mixing angle $\theta_{23}$ lies 
in the lower ($\theta_{23}<\pi/4$) 
or higher ($\theta_{23}>\pi/4$) octant.
These unknowns, which are essential to a 
complete understanding of neutrino mass and mixing,
can be probed by long-baseline neutrino experiments.

The MINOS long-baseline experiment~\cite{Michael:2008bc} 
has published measurements of oscillations using accelerator and atmospheric neutrinos and antineutrinos.
The oscillations observed by MINOS are driven by the larger mass-squared difference $\Delta m^{2}_{32}$; 
hence, many features of the data can be described by an
effective two-flavor model with a single mass-squared difference $\Delta m^{2}$
and mixing angle $\theta$. In this approximation, 
the $\nu_{\mu}$ and $\overline{\nu}_{\mu}$ survival probabilities are:

\begin{equation}
P(\nu_{\mu}\rightarrow\nu_{\mu}) \approx 1 - \sin^{2}2\theta\mbox{ }\sin^{2}\left(\frac{\Delta m^{2} L_{\nu}}{4E_{\nu}}\right),
\label{eqn:2flav}
\end{equation}

\noindent
where $L_{\nu}$ is the neutrino propagation distance and
$E_{\nu}$ is the neutrino energy.
A previous two-flavor analysis of $\nu_{\mu}$ and $\overline{\nu}_{\mu}$ disappearance
using the combined accelerator and atmospheric data from MINOS yielded
$|\Delta m^{2}| = 2.41^{+0.09}_{-0.10} \times 10^{-3}\mbox{\,eV}^{2}$
and $\sin^{2}2\theta = 0.950^{+0.035}_{-0.036}$~\cite{Adamson:2013whj}.
The statistical weight of the data now enables MINOS to constrain
the full three-flavor model of $\nu_{\mu}$ and $\overline{\nu}_{\mu}$ disappearance.
The uncertainty on $\Delta m^{2}$  is approaching the size 
of the smaller mass-squared difference, $\Delta m^{2}_{21}$, 
which is neglected in the two-flavor model.
Moreover, the precise knowledge of $\theta_{13}$ enables
an analysis of the data based on the full set of mixing parameters.
In this paper we present the three-flavor analysis of the combined MINOS data.

In the three-flavor framework, 
the oscillations are driven by two mass-squared
differences $\Delta m^{2}_{32}$ and $\Delta m^{2}_{31}$,
where $\Delta m^{2}_{31} = \Delta m^{2}_{32} + \Delta m^{2}_{21}$.
The interference between the resulting two oscillation frequencies 
leads to terms in the oscillation probabilities that depend 
on all the mixing parameters.
The leading-order $\nu_{\mu}$ and $\overline{\nu}_{\mu}$ survival probabilities 
in vacuum take the same form as the two-flavor approximation in Eq.~(\ref{eqn:2flav}), 
with the effective parameters given by~\cite{Nunokawa:2005nx}:

\begin{equation}
\begin{split}
\sin^{2}2\theta =& 4\sin^{2}\theta_{23}\cos^{2}\theta_{13}(1-\sin^{2}\theta_{23}\cos^{2}\theta_{13}),\\
\Delta m^{2} =& \Delta m^{2}_{32} + \Delta m^{2}_{21} \sin^{2}\theta_{12}\\
              &+ \Delta m^{2}_{21} \cos\delta_{CP}\sin\theta_{13}\tan\theta_{23}\sin2\theta_{12}.
\end{split}
\label{eqn:3flavDis}
\end{equation}

\noindent
The exact symmetries of the two-flavor model under $\theta \rightarrow \pi/2 - \theta$
and $\Delta m^{2} \rightarrow -\Delta m^{2}$ lead to approximate degeneracies 
in the octant of $\theta_{23}$ and mass hierarchy in the three-flavor formalism.

For neutrinos traveling through matter, the propagation eigenstates
are modified by the MSW effect~\cite{Wolfenstein:1977ue,*Mikheev:1986gs}.
In this case, the mixing angle $\theta_{13}$ is replaced by a 
modified version, $\theta_{M}$, given by~\cite{Giunti:1997fx}:

\begin{equation}
\sin^{2}2\theta_{M} = \frac{\sin^{2}2\theta_{13}}{\sin^{2}2\theta_{13}+(A-\cos2\theta_{13})^{2}}.
\label{eqn:3flavMatter}
\end{equation}

\noindent
The size of the matter effect is determined by the parameter
$A \equiv \pm 2\sqrt 2 G_{F} n_{e} E_{\nu} / \Delta m^{2}_{31}$,
where $G_{F}$ is the Fermi weak coupling constant,
$n_{e}$ is the density of electrons
and the sign of $A$ is positive (negative) for neutrinos (antineutrinos).
Equation~(\ref{eqn:3flavMatter}) shows that
$\sin^{2}2\theta_{M}$ is maximal at $A=\cos2\theta_{13}$.
This condition leads to the resonant enhancement of
$\nu_{\mu} \leftrightarrow \nu_{e}$ oscillations,
which can significantly alter the magnitude of $\nu_{\mu}$ disappearance.
The effect is present for neutrinos in the normal hierarchy 
and for antineutrinos in the inverted hierarchy.
An MSW resonance is predicted to occur in multi-GeV, upward-going atmospheric neutrinos,
which travel through the earth's mantle~\cite{PalomaresRuiz:2004tk}.
MINOS is the first experiment to probe this resonance by 
measuring $\nu_{\mu}$ and $\overline\nu_{\mu}$ interactions separately
with atmospheric neutrinos,
yielding sensitivity to the mass hierarchy and $\theta_{23}$ octant.

MINOS~\cite{Adamson:2013ue} has
previously reported measurements of $\nu_{e}$ and $\overline{\nu}_{e}$ appearance
in accelerator $\nu_{\mu}$ and $\overline{\nu}_{\mu}$ beams.
Measurements of $\nu_{\mu} \rightarrow \nu_{e}$ appearance in accelerator neutrinos
have also been published by T2K~\cite{Abe:2013hdq}.
Both results are based on three-flavor analyses.
For accelerator neutrinos, the $\nu_{\mu}\rightarrow\nu_{e}$ appearance probability in matter,
expanded to second order in $\alpha \equiv \Delta m^{2}_{21}/\Delta m^{2}_{31}$ $(\approx 0.03)$, 
is given by~\cite{Cervera:2000kp}:

\begin{equation}
\begin{split}
P(\nu_{\mu}\rightarrow\nu_{e}) \approx & \sin^{2}\theta_{23}\sin^{2}2\theta_{13}\frac{\sin^{2}\Delta(1-A)}{(1-A)^{2}}\\
                                       &+ \alpha \tilde{J} \cos(\Delta \pm \delta_{CP}) \frac{\sin\Delta A}{A}\frac{\sin\Delta(1-A)}{(1-A)}\\
                                       &+ \alpha^{2}\cos^{2}\theta_{23}\sin^{2}2\theta_{12}\frac{\sin^{2}\Delta A}{A^{2}}.
\end{split}
\label{eqn:3flavApp}
\end{equation}

\noindent
In this expression, $\tilde{J} \equiv \cos\theta_{13}\sin2\theta_{13}\sin2\theta_{12}\sin2\theta_{23}$,
$\Delta \equiv \Delta m^{2}_{31} L_{\nu} / 4 E_{\nu}$
and the plus (minus) sign applies to neutrinos (antineutrinos).
The first term in Eq.~(\ref{eqn:3flavApp}) is proportional 
to $\sin^{2}\theta_{23}$ and breaks the $\theta_{23}$ octant degeneracy.
In addition, the dependence on $A$ is sensitive to the mass hierarchy
and the second term in the expansion is sensitive to CP violation.
In this paper, we strengthen the constraints on $\delta_{CP}$, 
the $\theta_{23}$ octant and the mass hierarchy obtained from 
the MINOS appearance data~\cite{Adamson:2013ue} 
by combining the complete MINOS disappearance and appearance data
and by exploiting the improved precision on $\theta_{13}$ 
from reactor experiments.

In the MINOS experiment, 
the accelerator neutrinos are produced by 
the NuMI facility~\cite{Anderson:1998zza},
located at the Fermi National Accelerator Laboratory.
The complete MINOS accelerator neutrino data set comprises exposures 
of $10.71\times10^{20}$ protons-on-target (POT) 
using a \mbox{$\nu_{\mu}$-dominated} beam and $3.36\times10^{20}$\,POT 
using a \mbox{$\overline{\nu}_{\mu}$-enhanced} beam~\cite{Adamson:2013whj}.
These data were acquired in the ``low~energy'' 
NuMI beam configuration~\cite{Anderson:1998zza},
where the neutrino event energy peaks at 3\,GeV.
The~spectrum and composition of the beam are measured
using two steel-scintillator tracking detectors with
toroidal magnetic fields.
The Near and Far detectors are located  
1.04\,km and 735\,km downstream of the production target,
respectively. 
The 5.4\,kton Far Detector is installed 705\,m (2070\,m water-equivalent) underground
in the Soudan Underground Laboratory and is
equipped with a scintillator veto shield for rejection of
cosmic-ray muons. These features have enabled \mbox{MINOS} to collect
37.88\,~kton-years of atmospheric neutrino data~\cite{Adamson:2012gt}.

The oscillation analysis uses charged-current (CC) 
interactions of both muon and electron neutrinos.
These events are distinguished from neutral-current (NC) 
backgrounds by the presence of a muon track or electromagnetic shower, 
respectively. The events also typically contain shower activity from the hadronic recoil system.
The selection of accelerator $\nu_{\mu}$~CC and $\overline{\nu}_{\mu}$~CC events
is based on a multivariate $k$-Nearest-Neighbor classification algorithm
using a set of input variables characterizing the
topology and energy deposition of muon tracks~\cite{Ospanov:2008zza}.
The selected events are separated into contained-vertex neutrinos,
with reconstructed interaction positions inside
the fiducial volume of the detectors,
and non-fiducial muons, 
in which the neutrino interactions occur outside the fiducial volume 
or in the surrounding rock.
The contained-vertex events are further divided into 
candidate $\nu_{\mu}$ and $\overline{\nu}_{\mu}$ interactions
based on the curvature of their muon tracks.
In the oscillation fit, the events are binned as a 
function of reconstructed neutrino energy.
For contained-vertex events, this is taken as the sum
of the muon and hadronic shower energy measurements;
for~non-fiducial muons, the muon energy alone is used as
the neutrino energy estimator.
To improve the sensitivity to oscillations,
the contained-vertex $\nu_{\mu}$ events from the $\nu_{\mu}$-dominated beam
are also binned according to their calculated 
energy resolution~\cite{Mitchell:2011axi,Coleman:2011ita,Backhouse:2011zz}.
The predicted energy spectra in the Far Detector 
are derived from the observed data in the Near Detector
using a beam transfer matrix~\cite{Adamson:2007gu}.

\begin{figure*}[t]
   \includegraphics[width=\textwidth]{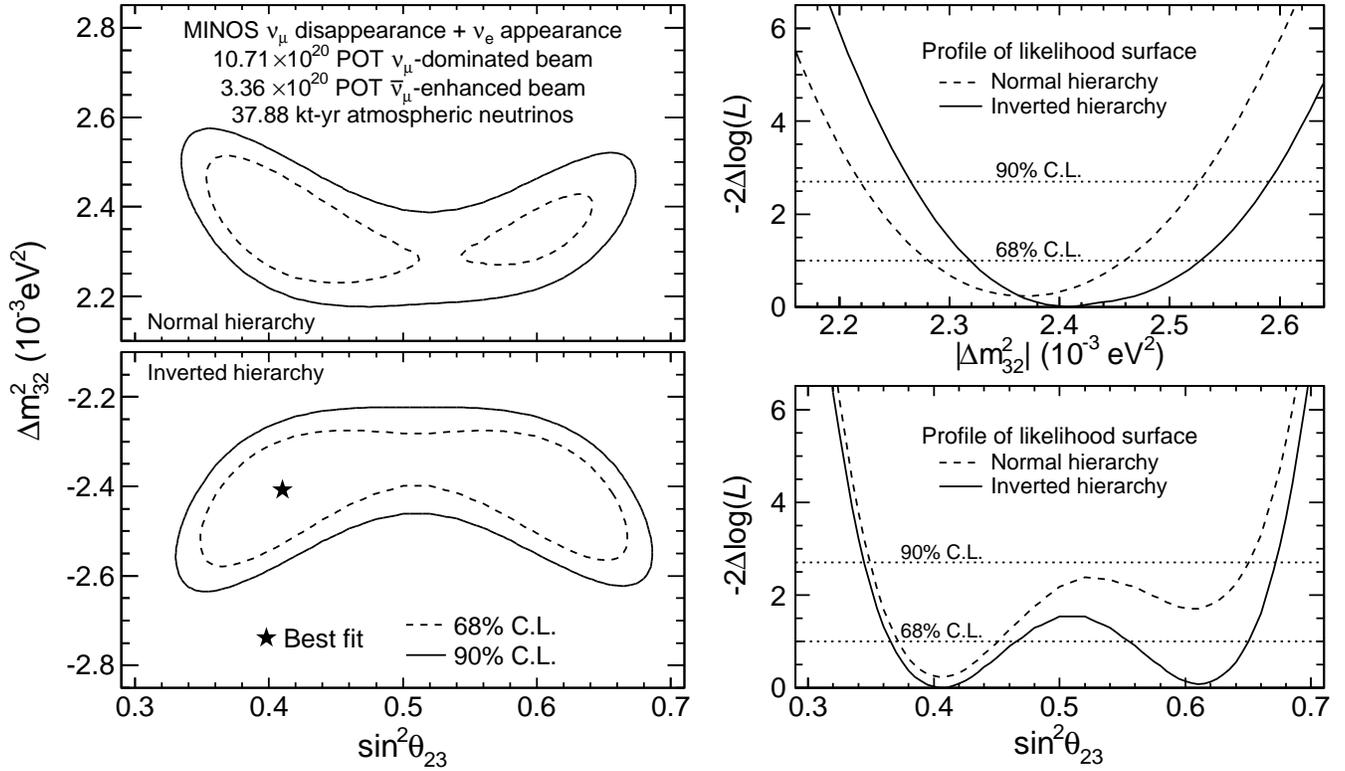}
   \caption{\label{contours_and_profiles}
     The left panels show the 68\% and 90\% confidence limits (C.L.)
     on $\Delta m_{32}^{2}$ and $\sin^{2}\theta_{23}$ for the normal hierarchy (top) 
     and inverted hierarchy (bottom).
     At each point in this parameter space, the likelihood function 
     is maximized with respect to $\sin^{2}\theta_{13}$, $\delta_{CP}$ and all of the systematic parameters.
     The $-2\Delta\log(\mathcal{L})$ surface is calculated relative to 
     the overall best fit, which is indicated by the star.
     The right panels show the 1D likelihood profiles as a function 
     of $\Delta m_{32}^{2}$ and $\sin^{2}\theta_{23}$ for each hierarchy. 
     The horizontal dotted lines indicate the 68\% and 90\% C.L.} 
 \end{figure*}

The selection of accelerator $\nu_{e}$~CC and $\overline{\nu}_{e}$~CC events
is based on a library-event-matching (LEM) algorithm
that performs hit-by-hit comparisons of
contained-vertex shower-like events with a large library 
of simulated neutrino interactions~\cite{OchoaRicoux:2009zz,Toner:2011xaa,Schreckenberger:2013gfa}.
The events are required to have reconstructed energies in the range $1-8$~GeV,
where most of the $\nu_{e}$ and $\overline{\nu}_{e}$ appearance
is predicted to occur.
The 50~best-matching events from the library are
used to calculate a set of classification
variables that are combined into a single discriminant
using an artificial neural network.
The selection does not discriminate between
$\nu_{e}$ and $\overline{\nu}_{e}$ interactions.
The selected events are binned as a function
of the reconstructed energy and LEM discriminant.
The background contributions from NC, $\nu_{\mu}$~CC and $\overline{\nu}_{\mu}$~CC interactions,
and intrinsic $\nu_{e}$~CC and $\overline{\nu}_{e}$~CC interactions from the beam,
are determined using samples of Near Detector data
collected in different beam configurations.
The backgrounds in the Far Detector are calculated
from these Near Detector components~\cite{Coelho:Thesis}.
The rates of appearance in the Far Detector are derived from the
$\nu_{\mu}$ CC and $\overline{\nu}_{\mu}$ CC spectra measured 
in the Near Detector~\cite{Adamson:2013ue}.

Atmospheric neutrinos are separated from the cosmic-ray muon background
using selection criteria that identify either a contained-vertex interaction
or an upward-going or horizontal muon track~\cite{Chapman:Thesis,Speakman:Thesis}.
For contained-vertex events, the background is further reduced 
by checking for associated energy deposits in the veto shield.
The event selection yields samples of contained-vertex and
non-fiducial muons, which are each separated into candidate
$\nu_{\mu}$~CC and $\overline{\nu}_{\mu}$~CC interactions.
These samples of muons are binned as a function of 
$\log_{10}(E)$ and $\cos\theta_{z}$,
where $E$ is the reconstructed energy of the event in GeV
and $\theta_{z}$ is the zenith angle of the muon track.
This two-dimensional binning scheme enhances the sensitivity to the MSW resonance.
The results remain in close agreement with the two-flavor
analysis of $\nu_{\mu}$ and $\overline{\nu}_{\mu}$ disappearance,
in which these data were binned as a function of $\log_{10}(L/E)$~\cite{Adamson:2013whj}.
A sample of contained-vertex showers is also selected from the data,
composed mainly of NC, $\nu_{e}$~CC and $\overline{\nu}_{e}$~CC interactions.
These events are grouped into a single bin,
since they have negligible sensitivity to oscillations but 
constrain the overall flux normalization.
The predicted event rates in each selected sample are calculated
from a Monte Carlo simulation of atmospheric neutrino
interactions in the Far Detector~\cite{Barr:2004br, Adamson:2012gt}.
The cosmic-ray muon backgrounds are obtained from the
observed data by reweighting the events tagged 
by the veto shield according to the measured shield inefficiency~\cite{Chapman:Thesis}.

\begin{table*}[t]
 \begin{ruledtabular}
 \begin{tabular}{ccccccc}
   Mass hierarchy & $\theta_{23}$ octant & $\Delta m_{32}^{2}\mbox{ }/\mbox{ }10^{-3}\mbox{eV}^{2}$ & $\sin^{2}\theta_{23}$ & $\sin^{2}\theta_{13}$ & $\delta_{CP} / \pi$ & $-2\Delta\log(\mathcal{L})$ \\[+0.025cm] 
 \hline \\[-0.25cm]
        $\Delta m^{2}_{32}<0$ & $\theta_{23}<\pi/4$ & $-2.41$ & $0.41$ & $0.0243$ & $0.62$ & $0$    \\
        $\Delta m^{2}_{32}<0$ & $\theta_{23}>\pi/4$ & $-2.41$ & $0.61$ & $0.0241$ & $0.37$ & $0.09$ \\
        $\Delta m^{2}_{32}>0$ & $\theta_{23}<\pi/4$ & $+2.37$ & $0.41$ & $0.0242$ & $0.44$ & $0.23$ \\
        $\Delta m^{2}_{32}>0$ & $\theta_{23}>\pi/4$ & $+2.35$ & $0.61$ & $0.0238$ & $0.62$ & $1.74$ \\
 \end{tabular}
 \end{ruledtabular}
 \caption{\label{best_fit_parameters}
    The best-fit oscillation parameters obtained from this analysis for each combination of 
    mass hierarchy and $\theta_{23}$ octant. Also listed are the $-2\Delta\log(\mathcal{L})$ values 
    for each of the four combinations, calculated relative to the overall best-fit point.} 
 \end{table*}

For all the data samples,
the predicted event spectra in the Far Detector are reweighted
to account for oscillations, and the backgrounds from 
$\nu_{\tau}$ and $\overline{\nu}_{\tau}$ appearance are included.
The oscillation probabilities 
are calculated directly from the PMNS matrix
using algorithms optimized for computational efficiency~\cite{Kopp:2006wp}.
The probabilities account for the propagation of neutrinos through the earth.
For accelerator neutrinos, a constant electron density of $1.36$\,mol\,cm$^{-3}$ is assumed.
For atmospheric neutrinos, 
the earth is modeled by four layers of constant electron density: 
\mbox{an inner core ($r<1220$\,km, $n_{e}=6.05$\,mol\,cm$^{-3}$)};
an outer core ($1220 < r < 3470$\,km, $n_{e}=5.15$\,mol\,cm$^{-3}$);
the mantle ($3470 < r < 6336$\,km, $n_{e}=2.25$\,mol\,cm$^{-3}$);
and the crust ($r > 6336$\,km, $n_{e}=1.45$\,mol\,cm$^{-3}$).
This four-layer approximation reflects the principal structures 
of the PREM model~\cite{Dziewonski:1981xy}.
Comparisons with a more detailed 42-layer model 
yield similar oscillation results.

The oscillation parameters are determined by applying 
a maximum likelihood fit to the data.
The parameters $\Delta m^{2}_{32}$, $\sin^{2}\theta_{23}$, 
$\sin^{2}\theta_{13}$ and $\delta_{CP}$ are varied
in the fit. 
The mixing angle $\theta_{13}$ 
is subject to an external constraint of
$\sin^{2}\theta_{13}=0.0242\pm0.0025$, 
based on a weighted average of the published results from
the Daya Bay~\cite{An:2012bu}, 
RENO~\cite{Ahn:2012nd}
and Double Chooz~\cite{Abe:2012tg}
reactor experiments.
This constraint is incorporated into the fit by adding 
a Gaussian penalty term to the likelihood function. 
The fit uses fixed values of 
$\Delta m_{21}^{2}=7.54\times10^{-5}\mbox{\,eV}^{2}$ 
and $\sin^{2}\theta_{12}=0.307$~\cite{Fogli:2012ua}.
The impact of these two parameters is evaluated 
by shifting them in the fit according to their uncertainties;
the resulting shifts in the fitted values of
$\Delta m_{32}^{2}$ and $\sin^{2}\theta_{23}$ are found to be negligibly small.
The likelihood function contains 32~nuisance parameters, 
with accompanying penalty terms, that account for the 
major systematic uncertainties in the simulation of the data~\cite{Adamson:2011ig,Adamson:2012gt,Toner:2011xaa}.
The fit proceeds by summing the separate likelihood contributions
from the $\nu_{\mu}$ disappearance~\cite{Adamson:2013whj}
and $\nu_{e}$ appearance~\cite{Adamson:2013ue} data sets,
taking their systematic parameters to be uncorrelated.

Figure~\ref{contours_and_profiles} shows the 2D confidence limits 
on $\Delta m_{32}^{2}$ and $\sin^{2}\theta_{23}$, obtained by
maximizing the likelihood function at each point in this
parameter space with respect to $\sin^{2}\theta_{13}$, $\delta_{CP}$ 
and all of the systematic parameters.
Also shown are the corresponding 1D likelihood profiles 
as a function of $\Delta m_{32}^{2}$ and $\sin^{2}\theta_{23}$.
The 68\% (90\%) confidence limits (C.L.) on these parameters
are calculated by taking the range of negative log-likelihood values with
$-2\Delta\log(\mathcal{L})<1.00$ ($2.71$) relative to the overall best fit. 
This yields $|\Delta m_{32}^{2}|=[2.28-2.46]\times10^{-3}\,\mbox{eV}^{2}$ (68\% C.L.)
and $\sin^{2}\theta_{23}=0.35-0.65$ (90\% C.L.)
in the normal hierarchy;
and $|\Delta m_{32}^{2}|=[2.32-2.53]\times10^{-3}\,\mbox{eV}^{2}$ (68\% C.L.)
and $\sin^{2}\theta_{23}=0.34-0.67$ (90\% C.L.)
in the inverted hierarchy.
The data disfavor maximal mixing ($\theta_{23}=\pi/4$)
by $-2\Delta\log(\mathcal{L})=1.54$.
The measurements of $\Delta m_{32}^{2}$ are the most precise 
that have been reported to date.

\begin{figure}
   \includegraphics[width=\columnwidth]{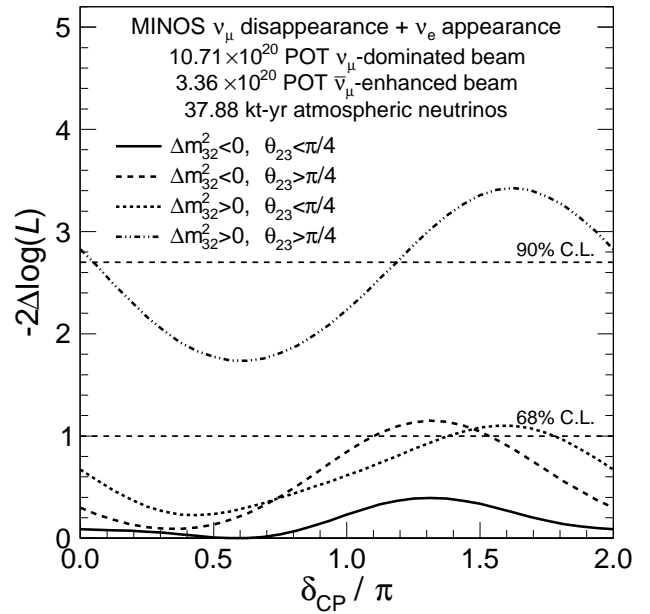}
   \caption{\label{delta_profile}
     The 1D likelihood profile as a function of $\delta_{CP}$ for each combination 
     of mass hierarchy and $\theta_{23}$ octant.
     For each value of $\delta_{CP}$, the likelihood function is 
     maximized with respect to $\sin^{2}\theta_{13}$, $\sin^{2}\theta_{23}$, $\Delta m_{32}^{2}$ 
     and all of the systematic parameters. 
     The horizontal dashed lines indicate the 68\% and 90\% confidence limits.} 
 \end{figure}

The data also constrain
$\delta_{CP}$, the $\theta_{23}$ octant degeneracy and the mass hierarchy.
Table~\ref{best_fit_parameters} lists the best-fit oscillation
parameters for each combination of octant and mass hierarchy,
and the differences in negative log-likelihood relative 
to the overall best fit.
Assuming $\theta_{23} > \pi/4$ ($\theta_{23} < \pi/4$), 
the data prefer the inverted hierarchy by $-2\Delta\log(\mathcal{L})=1.65\mbox{ }(0.23)$.
The combination of normal hierarchy and higher octant is
disfavored by $1.74$ units of $-2\Delta\log(\mathcal{L})$,
strengthening the previous constraints from 
$\nu_{e}$ and $\overline{\nu}_{e}$ appearance~\cite{Adamson:2013ue}.
Figure~\ref{delta_profile} shows the 1D likelihood profile as 
a function of $\delta_{CP}$ for each of the four possible combinations.
The data disfavor 36\% (11\%) of the parameter space defined by
$\delta_{CP}$, the $\theta_{23}$ octant,
and the mass hierarchy at 68\% (90\%) C.L.

In summary, we have presented the first combined analysis of $\nu_{\mu}$ 
disappearance and $\nu_{e}$ appearance data by a long-baseline neutrino experiment.
The results are based on the complete set of MINOS accelerator and atmospheric neutrino data. 
A combined analysis of these data sets yields precision measurements 
of $\Delta m^{2}_{32}$ and $\sin^{2}\theta_{23}$,
along with new constraints on the three-parameter space defined by $\delta_{CP}$,
the $\theta_{23}$ octant, and the mass hierarchy.

This work was supported by the U.S. DOE; the United Kingdom STFC;
the U.S. NSF; the State and University of Minnesota;
Brazil's FAPESP, CNPq and CAPES.
We are grateful to the Minnesota Department of Natural Resources
and the personnel of the Soudan Laboratory and Fermilab
for their contributions to the experiment.
We thank the Texas Advanced Computing Center at 
The University of Texas at Austin for the provision
of computing resources.

\bibliography{three_flavour_paper}

\end{document}

%% file: ThreeFlavor-Nov13-authors.tex
%
%
%
%
%
%
%

\newcommand{\Berkeley}{Lawrence Berkeley National Laboratory, Berkeley, California, 94720 USA}
\newcommand{\Cambridge}{Cavendish Laboratory, University of Cambridge, Madingley Road, Cambridge CB3 0HE, United Kingdom}
\newcommand{\Cincinnati}{Department of Physics, University of Cincinnati, Cincinnati, Ohio 45221, USA}
\newcommand{\FNAL}{Fermi National Accelerator Laboratory, Batavia, Illinois 60510, USA}
\newcommand{\RAL}{Rutherford Appleton Laboratory, Science and Technologies
Facilities Council, Didcot, OX11 0QX, United Kingdom}
\newcommand{\UCL}{Department of Physics and Astronomy, University College London, Gower Street, London WC1E 6BT, United Kingdom}
\newcommand{\Caltech}{Lauritsen Laboratory, California Institute of Technology, Pasadena, California 91125, USA}
\newcommand{\Alabama}{Department of Physics and Astronomy, University of Alabama, Tuscaloosa, Alabama 35487, USA}
\newcommand{\ANL}{Argonne National Laboratory, Argonne, Illinois 60439, USA}
\newcommand{\Athens}{Department of Physics, University of Athens, GR-15771 Athens, Greece}
\newcommand{\NTUAthens}{Department of Physics, National Tech. University of Athens, GR-15780 Athens, Greece}
\newcommand{\Benedictine}{Physics Department, Benedictine University, Lisle, Illinois 60532, USA}
\newcommand{\BNL}{Brookhaven National Laboratory, Upton, New York 11973, USA}
\newcommand{\CdF}{APC -- Universit\'{e} Paris 7 Denis Diderot, 10, rue Alice Domon et L\'{e}onie Duquet, F-75205 Paris Cedex 13, France}
\newcommand{\Cleveland}{Cleveland Clinic, Cleveland, Ohio 44195, USA}
\newcommand{\Delhi}{Department of Physics \& Astrophysics, University of Delhi, Delhi 110007, India}
\newcommand{\GEHealth}{GE Healthcare, Florence South Carolina 29501, USA}
\newcommand{\Harvard}{Department of Physics, Harvard University, Cambridge, Massachusetts 02138, USA}
\newcommand{\HolyCross}{Holy Cross College, Notre Dame, Indiana 46556, USA}
\newcommand{\Houston}{Department of Physics, University of Houston, Houston, Texas 77204, USA}
\newcommand{\IIT}{Department of Physics, Illinois Institute of Technology, Chicago, Illinois 60616, USA}
\newcommand{\Iowa}{Department of Physics and Astronomy, Iowa State University, Ames, Iowa 50011 USA}
\newcommand{\Indiana}{Indiana University, Bloomington, Indiana 47405, USA}
\newcommand{\ITEP}{High Energy Experimental Physics Department, ITEP, B. Cheremushkinskaya, 25, 117218 Moscow, Russia}
\newcommand{\JMU}{Physics Department, James Madison University, Harrisonburg, Virginia 22807, USA}
\newcommand{\LASL}{Nuclear Nonproliferation Division, Threat Reduction Directorate, Los Alamos National Laboratory, Los Alamos, New Mexico 87545, USA}
\newcommand{\Lebedev}{Nuclear Physics Department, Lebedev Physical Institute, Leninsky Prospect 53, 119991 Moscow, Russia}
\newcommand{\LLL}{Lawrence Livermore National Laboratory, Livermore, California 94550, USA}
\newcommand{\LosAlamos}{Los Alamos National Laboratory, Los Alamos, New Mexico 87545, USA}
\newcommand{\Manchester}{School of Physics and Astronomy, University of Manchester, Oxford Road, Manchester M13 9PL, United Kingdom}
\newcommand{\MIT}{Lincoln Laboratory, Massachusetts Institute of Technology, Lexington, Massachusetts 02420, USA}
\newcommand{\Minnesota}{University of Minnesota, Minneapolis, Minnesota 55455, USA}
\newcommand{\Crookston}{Math, Science and Technology Department, University of Minnesota -- Crookston, Crookston, Minnesota 56716, USA}
\newcommand{\Duluth}{Department of Physics, University of Minnesota Duluth, Duluth, Minnesota 55812, USA}
\newcommand{\Ohio}{Center for Cosmology and Astro Particle Physics, Ohio State University, Columbus, Ohio 43210 USA}
\newcommand{\Otterbein}{Otterbein College, Westerville, Ohio 43081, USA}
\newcommand{\Oxford}{Subdepartment of Particle Physics, University of Oxford, Oxford OX1 3RH, United Kingdom}
\newcommand{\PennState}{Department of Physics, Pennsylvania State University, State College, Pennsylvania 16802, USA}
\newcommand{\PennU}{Department of Physics and Astronomy, University of Pennsylvania, Philadelphia, Pennsylvania 19104, USA}
\newcommand{\Pittsburgh}{Department of Physics and Astronomy, University of Pittsburgh, Pittsburgh, Pennsylvania 15260, USA}
\newcommand{\IHEP}{Institute for High Energy Physics, Protvino, Moscow Region RU-140284, Russia}
\newcommand{\Rochester}{Department of Physics and Astronomy, University of Rochester, New York 14627 USA}
\newcommand{\RoyalH}{Physics Department, Royal Holloway, University of London, Egham, Surrey, TW20 0EX, United Kingdom}
\newcommand{\Carolina}{Department of Physics and Astronomy, University of South Carolina, Columbia, South Carolina 29208, USA}
\newcommand{\SLAC}{Stanford Linear Accelerator Center, Stanford, California 94309, USA}
\newcommand{\Stanford}{Department of Physics, Stanford University, Stanford, California 94305, USA}
\newcommand{\StJohnFisher}{Physics Department, St. John Fisher College, Rochester, New York 14618 USA}
\newcommand{\Sussex}{Department of Physics and Astronomy, University of Sussex, Falmer, Brighton BN1 9QH, United Kingdom}
\newcommand{\TexasAM}{Physics Department, Texas A\&M University, College Station, Texas 77843, USA}
\newcommand{\Texas}{Department of Physics, University of Texas at Austin, 1 University Station C1600, Austin, Texas 78712, USA}
\newcommand{\TechX}{Tech-X Corporation, Boulder, Colorado 80303, USA}
\newcommand{\Tufts}{Physics Department, Tufts University, Medford, Massachusetts 02155, USA}
\newcommand{\UNICAMP}{Universidade Estadual de Campinas, IFGW-UNICAMP, CP 6165, 13083-970, Campinas, SP, Brazil}
\newcommand{\UFG}{Instituto de F\'{i}sica, Universidade Federal de Goi\'{a}s, CP 131, 74001-970, Goi\^{a}nia, GO, Brazil}
\newcommand{\USP}{Instituto de F\'{i}sica, Universidade de S\~{a}o Paulo,  CP 66318, 05315-970, S\~{a}o Paulo, SP, Brazil}
\newcommand{\Warsaw}{Department of Physics, University of Warsaw, Ho\.{z}a 69, PL-00-681 Warsaw, Poland}
\newcommand{\Washington}{Physics Department, Western Washington University, Bellingham, Washington 98225, USA}
\newcommand{\WandM}{Department of Physics, College of William \& Mary, Williamsburg, Virginia 23187, USA}
\newcommand{\Wisconsin}{Physics Department, University of Wisconsin, Madison, Wisconsin 53706, USA}
\newcommand{\deceased}{Deceased.}

\affiliation{\ANL}
\affiliation{\Athens}
\affiliation{\BNL}
\affiliation{\Caltech}
\affiliation{\Cambridge}
\affiliation{\UNICAMP}
\affiliation{\Cincinnati}
\affiliation{\FNAL}
\affiliation{\UFG}
\affiliation{\Harvard}
\affiliation{\HolyCross}
\affiliation{\Houston}
\affiliation{\IIT}
\affiliation{\Indiana}
\affiliation{\Iowa}
\affiliation{\UCL}
\affiliation{\Manchester}
\affiliation{\Minnesota}
\affiliation{\Duluth}
\affiliation{\Otterbein}
\affiliation{\Oxford}
\affiliation{\Pittsburgh}
\affiliation{\RAL}
\affiliation{\USP}
\affiliation{\Carolina}
\affiliation{\Stanford}
\affiliation{\Sussex}
\affiliation{\TexasAM}
\affiliation{\Texas}
\affiliation{\Tufts}
\affiliation{\Warsaw}
\affiliation{\WandM}

\author{P.~Adamson}
\affiliation{\FNAL}


\author{I.~Anghel}
\affiliation{\Iowa}
\affiliation{\ANL}



\author{A.~Aurisano}
\affiliation{\Cincinnati}








\author{G.~Barr}
\affiliation{\Oxford}









\author{M.~Bishai}
\affiliation{\BNL}

\author{A.~Blake}
\affiliation{\Cambridge}


\author{G.~J.~Bock}
\affiliation{\FNAL}


\author{D.~Bogert}
\affiliation{\FNAL}




\author{S.~V.~Cao}
\affiliation{\Texas}

\author{C.~M.~Castromonte}
\affiliation{\UFG}



% ...

\collaboration{The MINOS Collaboration}
\noaffiliation

%
%
%
%